\documentclass[twocolumn]{aastex63}

\newcommand\aastex{AAS\TeX}
\newcommand\latex{La\TeX}

\shorttitle{Mach's Principle}
\shortauthors{Szapudi}
\graphicspath{{./}{figures/}}

\begin{document}

\title{Constraining Mach's principle with high precision astrometry}

\correspondingauthor{Istv\'an Szapudi}
\email{szapudi@ifa.hawaii.edu}

\author[0000-0003-2274-0301]{Istv\'an Szapudi}
\affiliation{Institute for Astronomy \\
University of Hawai'i at M$\bar{a}$noa \\ 
2680 Woodlawn Drive\\ 
Honolulu, HI 96822, USA}





\begin{abstract}

The analyses of high precision astrometric surveys, such as Gaia, implicitly assume a modern version of Mach's Principle: the local inertial frame of our Solar System should be non-rotating in the frame of distant quasars. On the contrary,  Einstein's General Relativity allows a rotating universe. Thus, relaxing the assumption of Mach's Principle will allow placing a constraint on a class of rotating cosmologies by comparing high precision astrometry of quasars with well-measured solar system orbits. Constraining global rotation will test General Relativity, inflation, and the isotropy of cosmological initial conditions. 

{\it Essay written for the Gravity Research Foundation 2021 Awards for Essays on Gravitation.}

\end{abstract}



\section{Introduction}

\label{sec:intro}

According to Mach's Principle (MP, attributed to Ernst Mach and Bishop Berkeley) rotation cannot be observed without distant reference objects, thus inertial frames are at rest with respect to and somehow a consequence of distant stars.

While MP motivated Einstein's General Relativity (GR), it is a non-Machian local field theory with no action at a distance. Indeed, the rotation of the Milky Way breaks MP in the original sense. The modern view (aside from trivially replacing galactic stars with distant quasars and galaxies) is that inertial frames are local and not caused by distant objects. The exception is frame dragging \citep[See][for the history of the Lense-Thirring effect]{Pfister2007} on non-cosmological scales. A gyroscope rotating (or Sagnac effect) in space with respect to distant quasars signals that the universe rotates. 

For a cosmologist, the non-rotating FRW solution appears unnatural, analogous to non-rotating Schwarzschild blackholes vs. rotating Kerr blackholes.  The crucial difference is that blackholes form through collapse while the universe expands. Collapse increases angular momentum through momentum conservation, while cosmological angular momentum (and any vector perturbation) decreases with $1/a$, where $a$ is the expansion factor of the universe. Thus we would expect only a small amount of residual rotation in most theories. In particular, in standard inflationary theories the expansion of order $a \simeq e^{60}$ could render any residual rotation unmeasurable. Nevertheless, some non-standard inflationary models, \citep[e.g., the Einstein-aether theory][]{EinsteinAether2011} seed vector perturbations. In conclusion, any observed rotation of our inertial frame (or the lack thereof) constrains inflation, the isotropy of cosmological initial conditions, and GR.

Rotating models have been considered early on by \cite{godel}, and later \cite[e.g.,][]{silk,Hawking1969,CollinsHawking1973}. While the original G\"odel model is pathological, there are viable generalizations; for a review see \cite{obukhov}. These models display a global rotation while still preserving a perfectly uniform Cosmic Microwave Background \citep[CMB, e.g.,][]{obukhov}. 
Note that vector perturbations to homogeneous FRW and related Bianchi models belong to an interesting but separate class of models \citep{BarrowEtal1985,SuChu2009}, tightly constrained from the CMB \citep{McEwenEtal2013,SaadehEtal2016a,SaadehEtal2016b}. Our focus is on cosmologies still allowed by CMB constraints.

Gaia used distant quasars to determine the acceleration of the solar system at the precision of $0.35 \mu as/y$, \citep[][]{GaiaAcceleration}, with the expectation of ``well below $0.2 \mu as/y$ after the final data release''. {\bf However, their analyses implicitly use MP.} We demonstrate next that high precision astrometry combined with solar system observations and accurate dynamical modeling will constrain global rotation \citep[see also][for a constraint of $0.1 mas/y$ based on solar system data]{Clemence1957}.

\section{The effect of global rotation on solar system orbits}
\label{section:rotation}

To first order, our Solar system can be replaced with the Sun and Jupiter orbiting at 5AU, since the latter contains approximately $2/3$ of the total mass in planets. Thus, initially, let's assume solar system with a central star and a single planet. In addition, assume that the planet is far enough from the central star that general relativistic corrections are negligible, and that the local inertial frame rotates with respect to the frame defined by distant quasars. The solution in a rotating frame is a standard Keplerian orbit in a rotating plane. We notate the projection of the global rotation $\omega$ perpendicular to the plane of the orbit as $\omega_\perp$ and in the plane as $\omega_\parallel = \omega - \omega_\perp$.

High precision astrometry of distant stars and a precision determination of the plane of the orbit is sufficient to determine $\omega_\parallel$. From $N$ measurement of the orbit, we can determine the plane of the orbit with $\sigma/\sqrt{N}$, $\sigma$ is the fundamental astrometric precision for the planet orbit (depends on the brightness, color, etc. of the planet) in $\mu as$. If in a year this becomes smaller than the astrometric precision of the quasar frame in $\sigma_Q$ in $\mu as/y$, then $\omega_\parallel$ is measurable.

For constraining $\omega_\perp$, note that a rotating coordinate system pegged to distant quasars cases deviations to Kepler's law. For a single planet on a circular orbit of radius $R$ and velocity $v$ and period $P = 2\pi R/v$ the observed velocity is
\begin{equation}
    \tilde v = v+ \omega_\perp R.
\end{equation}
The measured period is modified as
\begin{equation}
    \tilde P = \frac{2\pi}{v+\omega_\perp R} = \simeq P\left(1 - \frac{\omega_\perp R}{v}\right),
\end{equation}
where we keep terms linear with the small $\omega_\perp$. Kepler's law is modified as 
\begin{equation}
    \tilde P^2 = \frac{4\pi^2 R^3}{G M} \left(1 - \frac{2 R^{3/2}}{(GM)^{1/2}} \omega_\perp \right).
\end{equation}
Precise measurements of the mass $M$ of the central star, the radius of the orbit, and the period will pinpoint a rotating frame.
To appreciate the order of magnitudes, we can estimated the expected corrections to the period as 
\begin{equation}
    \frac{\Delta P}{P} \simeq 7.7 \times 10^{-13} P[y] \omega_\perp [\mu as /y].
    \label{eq:dpperp}
\end{equation}

The above toy model Newtonian solar system with one planet illustrates how to constrain a rotating frame of distant quasars. In practice, we need a general relativistic multi-body model of our solar system. Next we outline the needed corrections to the above simple ideas.

Angular momentum conserves in a Schwartzschild metric commonly used to model the gravitational field of the Sun, therefore the kinematic measurement of $\omega_\parallel$ does not pick up GR corrections.

On the contrary, Kepler's law is modified in a GR case \citep[e.g.,][]{BarkerEtal1986}. To first order such corrections depend on the Schwartzschild radius of the Sun $R_s \simeq 3\, km \simeq 2 \times 10^{-8} AU$, thus relativistic corrections are likely to be important for the accuracy needed to constrain at the level Eq.~\ref{eq:dpperp}. The PPN formalism \citep{Will2014} with additional parameters to describe rotation can account for perturbative GR effects.

The interaction between the planets has significant effect on their orbits. For instance the semi-major axis of Mars can change order of $10^{-4}$ due to the effect of Earth, and Jupiter. Such perturbations alter mainly the period and semi major axis, but in principle they can tilt the instantaneous plane of the orbit. Therefore they could affect the measurement of both $\omega_\parallel$ and $\omega_\perp$. 

Thus rotation of our rest frame with respect to the quasar frame will show up in local observations of planetary dynamics coupled with precision astrometry. In practice, detailed, general relativistic models including planetary perturbations are necessary in conjunction with high precision quasar astrometry. Nevertheless, the principle remains the same: {\bf orbital elements and dynamics constrain the local inertial frame, our Solar System acting as a giant gyroscope.}

\section{Gaia}
\label{section:gaia}
 At present, Gaia provides the most precise astrometry. Their Early Data Release 3 (Gaia EDR3) determined precise astrometry for about 1.6 million compact (QSO-like) sources, 1.2 million of which have the best-quality five-parameter astrometric solutions. These latter have astrometric precision $\simeq 450 mas/y$. Naively, a coherent motion on the sky could be thus detected at the level of $0.4\mu as/y$. This order of magnitude consideration is remarkablt close to the quoted error on the detected acceleration of the solar system $0.35\mu as/y$ \citep{GaiaAcceleration}, despite the recently identified spatial correlation of the parallaxes \citep{Zinn2021}. 

The Gaia analysis expands the proper motion of quasars into spheroidal and toroidal vector spherical harmonics (VSHs) according to \citep{VSH}. Rotation would appear as linear combination of $l=1$ toroidal VSHs. Gaia simulations predict an error of order $0.1 \mu as$  for constraining rotation, again, close to the order of magnitude estimates earlier.

The present Gaia analysis explicitly assumes that QSOs provide a fixed frame, i.e. they presuppose MP. Individual VSH coefficients are not presented in \citep{GaiaAcceleration}: the $l=1$ toroidal VSHs provide a check on systematic errors. They use only the glide coefficients derived from the spherical VSHs. {\bf Comparison of the measured VSH coefficients with solar system orbits will yield the first constraint on generalized G\"odel cosmologies.}

\section{Summary and Discussion}

While MP has several versions and interpretations in GR, we focus on whether our local inertial frame is rotating with respect to distant objects, QSOs and galaxies. We demonstrated how precision astrometry combined with accurate general relativistic multi-body model of the Solar system will constrain MP in this sense. By analogy of blackholes, a small rotation would be natural in GR, while a perfect synchronization with the background, MP, would be curious.  In standard inflationary models we expect that initial rotation would be subdued by expansion. Our proposed astrometric constraints will shed light on inflationary theories, the isotropy of cosmology and initial conditions, and GR. Next we propose a few additional tests for global rotation.


Any satellite, such as Gaia, is a free floating laboratory in the solar system. Therefore, up to orbital microgravity, its gyroscope determines the local inertial frame. Thus the ephemeris of a satellite together with gyroscopic (or Sagnac) data compared to QSO positions yield a constraint. At the same time, Gaia is close enough to other rotating bodies, such as the Sun, that the Lense-Thirring frame dragging could complicate interpretation.

Since angular momentum is conserved even in a GR setting, one could measure the total angular momentum of a few interacting Solar System bodies, such as Jupiter, Saturn, etc., to define a conserved plane. If the torque from any neglected bodies is negligible, the conserved reference plane constrains rotation as in Section~\ref{section:rotation}. 

Expansion slows any rotation by $1/a$, therefore higher redshift quasars should spin faster. Subdividing a sample into (photometric) redshift slices enhances the signal compared to the solar system, displays differential rotation. High resolution CMB observations in the quasar frame would show a factor of 1000 faster rotation compared to local quasars.

Finally, a similar argument works for galactic orbits of stars. In the (perhaps far) future, if sufficiently accurate measurements of stellar positions are available, stellar orbits in the quasar frame will yield additional constraints on rotation.

\section{Acknowledgements}
IS acknowledges support from the National Science Foundation (NSF) award 1616974 and thanks Eric Baxter, Nader Haghighipour, Dan Huber, and Jeremy Sakstein for useful comments and suggestions.

\bibliography{sample63}{}
\bibliographystyle{aasjournal}



\end{document}